\documentclass[sigconf]{acmart}

\AtBeginDocument{%
  \providecommand\BibTeX{{%
    \normalfont B\kern-0.5em{\scshape i\kern-0.25em b}\kern-0.8em\TeX}}}

\setcopyright{acmcopyright}
\copyrightyear{2020}
\acmYear{2020}
\setcopyright{acmcopyright}\acmConference[SIGIR '20]{Proceedings of the 43rd International ACM SIGIR Conference on Research and Development in Information Retrieval}{July 25--30, 2020}{Virtual Event, China}
\acmBooktitle{Proceedings of the 43rd International ACM SIGIR Conference on Research and Development in Information Retrieval (SIGIR '20), July 25--30, 2020, Virtual Event, China}
\acmPrice{15.00}
\acmDOI{10.1145/3397271.3401253}
\acmISBN{978-1-4503-8016-4/20/07}

\usepackage{amsmath,amssymb,amsfonts, amsthm}
\usepackage{algorithmic,algorithm}
\usepackage{graphicx}
\usepackage{subcaption}
\usepackage[belowskip=2pt,aboveskip=3pt]{caption}
\usepackage{textcomp}
\usepackage{url}
\usepackage{multirow}
\usepackage{array}
\usepackage{enumitem}

\renewcommand\abstractname{ABSTRACT}

\theoremstyle{definition}

\begin{document}
\fancyhead{}
\title{Alleviating the Inconsistency Problem of Applying Graph Neural Network to Fraud Detection}


\author{Zhiwei Liu, Yingtong Dou, Philip S. Yu}
\affiliation{%
  \institution{Department of Computer Science,\\University of Illinois at Chicago}
}
\email{{zliu213, ydou5, psyu}@uic.edu}
\author{Yutong Deng}
\affiliation{%
  \institution{School of Software, \\Beijing University of Posts and Telecommunications}
}
\email{buptdyt@bupt.edu.cn}
\author{Hao Peng}
\affiliation{%
  \institution{Beijing Advanced Innovation Center for Big Data and Brain Computing, Beihang University}
}
\email{penghao@act.buaa.edu.cn}


\begin{abstract}
Graph-based models have been widely used to fraud detection tasks. Owing to the development of Graph Neural Networks~(GNNs), recent works have proposed many GNN-based fraud detectors, which are based on either homogeneous or heterogeneous graphs. These works design some GNNs, aggregating neighborhood information to learn the node embeddings. The aggregation relies on the assumption that neighbors share similar context, features, and relations. However, the inconsistency problem incurred by fraudsters is hardly investigated, i.e., the context inconsistency, feature inconsistency, and relation inconsistency. In this paper, we introduce these inconsistencies and design a new GNN framework, $\mathsf{GraphConsis}$, to tackle the inconsistency problem: (1) for the context inconsistency, we propose to combine the context embeddings with node features; (2) for the feature inconsistency, we design a consistency score to filter the inconsistent neighbors and generate corresponding sampling probability; (3) for the relation inconsistency, we learn the relation attention weights associated with the sampled nodes. Empirical analyses demonstrate that the inconsistency problem is critical in fraud detection tasks. Extensive experiments show the effectiveness of $\mathsf{GraphConsis}$. We also released a GNN-based fraud detection toolbox with implementations of SOTA models. The code is available at \textcolor{blue}{\url{https://github.com/safe-graph/DGFraud}}.


\end{abstract}

\ccsdesc[500]{Security and privacy~Web application security}
\ccsdesc[500]{Computing methodologies~Neural networks}

\keywords{Graph Neural Networks; Fraud Detection; Inconsistency Problem}



\maketitle

\section{INTRODUCTION}


There are various kinds of fraudulent activities on the Internet~\cite{jiang2016suspicious}, e.g., fraudsters disguise as regular users to post fake reviews~\cite{kaghazgaran2019wide} and commit download fraud~\cite{dou2019uncovering}.
By modeling entities as nodes and the corresponding interactions between entities as edges~\cite{peng2019fine}, we can design graph-based algorithms to detect suspicious patterns and therefore spot the fraudsters. Along with the development of Graph Neural Networks~(GNNs)~\cite{kipf2016semi,velivckovic2017graph,hamilton2017inductive}, previous endeavors propose many GNN-based fraud detection frameworks~\cite{wang2019fdgars, li2019spam,zhong2020financial,liu2018heterogeneous,wang2019semi,zhang2019key}. 

Among those frameworks, \cite{wang2019fdgars, li2019spam} detect opinion fraud in the online review system, \cite{zhong2020financial, liu2018heterogeneous, wang2019semi} aim at financial fraud, and \cite{zhang2019key} targets cyber-criminal in online forums.
They proposed new GNNs upon either homogeneous~\cite{wang2019fdgars, li2019spam} or heterogeneous~\cite{liu2018heterogeneous, zhong2020financial, li2019spam, wang2019semi,zhang2019key} graphs. 
Regarding base GNN models, FdGars~\cite{wang2019fdgars} and GAS\cite{li2019spam} adopt GCN~\cite{kipf2016semi}, SemiGNN and Player2Vec~\cite{zhang2019key} adopt GAT~\cite{velivckovic2017graph}, and other works~\cite{liu2018heterogeneous, li2019spam, zhong2020financial} devise new aggregators to aggregate the neighborhood information.
Those GNN-based fraud detectors learn node representation iteratively and predict the node suspiciousness in an end-to-end and semi-supervised fashion.

\begin{figure*}
    \centering
   \includegraphics[width=0.81\textwidth]{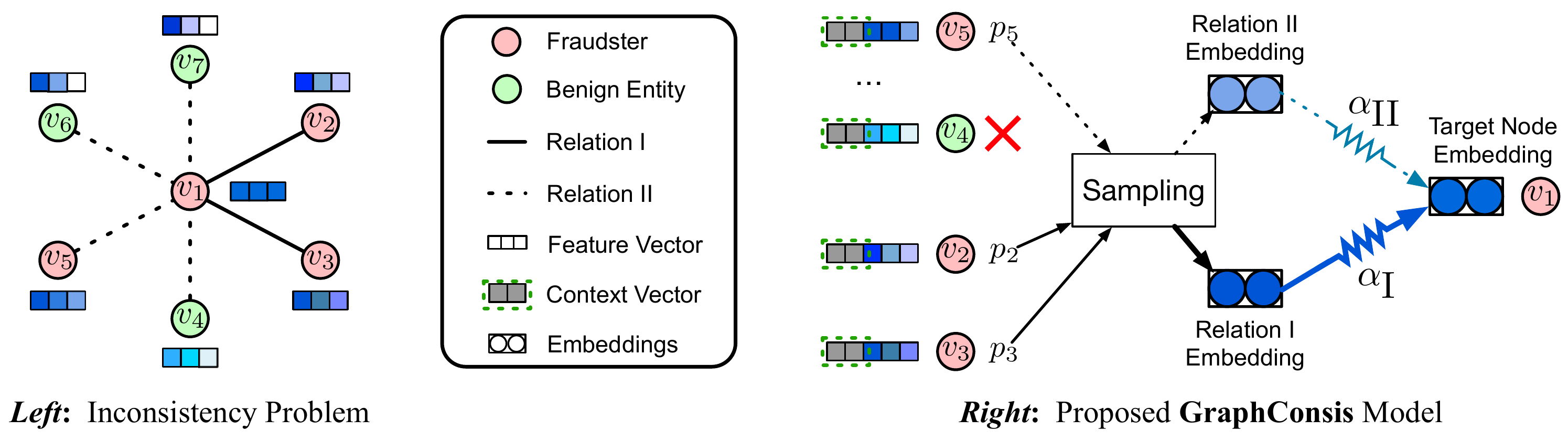}
    \caption{\small \textit{Left}: A toy example of a graph with two relations constructed on a fraud dataset, $v_2-v_7$ are neighbors of $v_1$. \textcolor{blue}{Context inconsistency}: fraudster $v_1$ can connect to many benign neighbors ($v_4,v_6,v_7$ in Relation II) to disguise itself. \textcolor{blue}{Feature inconsistency}: for $v_2$ and $v_3$ with the same relation to $v1$, their features may have great differences. \textcolor{blue}{Relation inconsistency}: for $v_1$, Relation I connects more similar neighbors than Relation II. \textit{Right}: To alleviate the inconsistency problem, we introduce three techniques. First, we propose to combine the context embeddings with feature vectors. Then, we calculate the consistency scores of neighbors to filter nodes and generate sampling probabilities. Finally, we aggregate the sampled neighbors with the attention mechanism over relation embeddings.}
    \label{fig:consis}
\end{figure*}

However, all existing methods ignore the inconsistency problem when designing a GNN model regarding the fraud detection task. The inconsistency problem is associated with the aggregation process of GNN models. The mechanism of aggregation is based on the assumption that the neighbors share similar features and labels~\cite{hou2020measure}. When the assumption diminishes, we can no longer aggregate neighborhood information to learn node embeddings. For example, as Figure~\ref{fig:consis} (\textit{Left}) shows, the inconsistency in fraud detection problem comes from three perspectives: 

\noindent (1) \textbf{Context inconsistency}. Smart fraudsters can connect themselves to regular entities as camouflage~\cite{kaghazgaran2018combating, sun2018adversarial}. Meanwhile, the amount of fraudsters is much less than that of regular entities. Directly aggregating neighbors by the GNNs can only help fraudulent entities aggregate the information from regular entities and thus prevent themselves from being spotted by fraud detectors. For example, in Figure~\ref{fig:consis} (\textit{Left}), the fraudster $v_1$ connects to $3$ benign entities under relation II.

\noindent (2) \textbf{Feature inconsistency}. Take the opinion fraud (e.g., spam reviews) detection problem as an example~\cite{li2019spam}. Assuming there are two reviews posted from the same user but regarding products in distinct categories, these two reviews have an edge since they share the same user. However, their review content (features) are far from each other, as they are associated with different products. Direct aggregation makes the GNN hard to distinguish the unique semantic characteristics of reviews and finally affects its ability to detect spam reviews. For example, in Figure~\ref{fig:consis} (\textit{Left}), we can observe that the feature of node $v_1$ is inconsistent with that of nodes $v_4$, $v_6$, and $v_7$. 

\noindent (3) \textbf{Relation inconsistency}. Since entities are connected with multiple types of relations, equally treating all relations results in a relation inconsistency problem. For example, two reviews may either be connected by the same user or the same product, which are respectively \textit{common-user} relation and \textit{common-product} relation. Assuming that one review is suspicious, the other one should have a higher suspiciousness if they are connected by \textit{common-user} relation, since fraudulent users tend to post more than one fraudulent reviews. For example, in Figure~\ref{fig:consis} (\textit{Left}), we find that under relation II, the fraudster $v_1$ is connected to two other fraudsters. However, under relation I, the fraudster is connected to only one fraudster but three benign entities. 

To tackle all above inconsistencies, we design a novel GNN framework, named as \textbf{GraphConsis}. The framework of $\mathsf{GraphConsis}$ is shown in Figure~\ref{fig:consis} (\textit{Right}). $\mathsf{GraphConsis}$ is built upon a heterogeneous graph with multiple relations. $\mathsf{GraphConsis}$ differs existing GNNs from the aggregation process. Instead of directly aggregating neighboring embeddings, we design three techniques to resolve three inconsistency problems simultaneously. Firstly, to handle the context inconsistency of neighbors, $\mathsf{GraphConsis}$ assigns each node a trainable context embedding, which is illustrated as the gray block aside nodes in Figure~\ref{fig:consis} (\textit{Right}). 
Secondly, to aggregate consistent neighbor embeddings, we design a new metric to measure the \textit{embedding consistency} between nodes. By incorporating the embedding consistency score into the aggregation process, we ignore the neighbors with a low consistency score (e.g. the node $v_4$ is dropped in Figure~\ref{fig:consis} (\textit{Right})) and generate the sampling probability. Last but not least, we learn relation attention weights associated with neighbors in order to alleviate the relation inconsistency problem. 

The contributions of this paper are:
\begin{itemize}

    \item To the best of our knowledge, we are the first work addressing the inconsistency problem in GNN models. 
    
    \item We empirically analyze three inconsistency problems regarding applying GNN models to fraud detection tasks. 
    
    \item We propose $\mathsf{GraphConsis}$ to tackle three inconsistency problems, which combines context embedding, neighborhood information measure, and relational attention. 
\end{itemize}

\section{PRELIMINARIES}
We detect fraud entities in the graph by using node representations. Hence, we first introduce node representation learning. A heterogeneous graph $G=\{V, \mathbf{X},  \{E_{r}\}|_{r=1}^{R}\big\}$, where $V$ denotes the nodes, $\mathbf{X}$ is the feature matrix of nodes, and $E_{r}$ denotes the edges w.r.t. the relation $r$. We have $R$ different types of relations. To represent the nodes as vectors, we need to learn a function $f:V\rightarrow \mathbb{R}^{d}$ that maps nodes to a $d$ dimensional space, where $d \ll |V|$.  The function $f$ should preserve both the structural information of the graph and the original feature information of the nodes. With the learned node embeddings, we can train a classifier $C:\mathbb{R}^{d}\rightarrow \{0,1\}$ to detect whether a given node is a fraudster, where $1$ denotes fraudster, and $0$ denotes benign entity. In this paper, we adopt the GNN framework to learn the node representation through neighbor aggregation. GNN framework can train the mapping function $f$ and the classifier $C$ simultaneously. We only need to input the graph and the labels of nodes to a GNN model. The general framework of a GNN model is:
\begin{equation}\label{eq:gnn_framework}
    \mathbf{h}_{v}^{(l)}= \mathbf{h}_{v}^{(l-1)} \oplus \operatorname{AGG}^{(l)}\left(\left\{\mathbf{h}_{v^{\prime}}^{(l-1)}: v^{\prime} \in \mathcal{N}_{v}\right\}\right),
\end{equation}
where $\mathbf{h}_{v}^{(l)}$ is the hidden embedding of $v$ at $l$-th layer, $\mathcal{N}_{v}$ denotes the neighbors of node $v$, and the $\operatorname{AGG}$ represents the aggregation function that maps the neighborhood information into a vector. Here, we use $\oplus$ to denote the combination of neighbor information and the center node information, it can be direct addition or concatenation then passed to a neural network. For the $\operatorname{AGG}$ function, we first assign a sampling probability to the neighboring nodes. Then we sample $Q$ nodes and average\footnote{Other pooling techniques can also be applied.} them as a vector. The calculation of probability is introduced later in Eq.~(\ref{eq:sampling prob}). Note that the framework of GNN is a $L$-layer structure, where $1\leq l \leq L$. At $l$-th layer, it aggregates the information from $l-1$-th layer. 



\section{PROPOSED MODEL}
\subsection{Context Embedding}
The aggregator combines the information of neighboring nodes according to Eq.~(\ref{eq:gnn_framework}). When $k=1$, the hidden embedding $\mathbf{h}^{(0)}_{v}$ is equivalent to the node feature. To tackle the context inconsistency problem, we introduce a trainable context embedding $\mathbf{c}_{v}$ for node $v$., instead of only using its feature vector $\mathbf{x}_{v}$. The first layer of the aggregator then becomes:
\begin{equation}\label{eq:gnn_framework}
    \mathbf{h}_{v}^{(1)}= \{\mathbf{x}_{v}\|\mathbf{c}_{v}\} \oplus \operatorname{AGG}^{(1)}\left(\left\{\mathbf{x}_{v^{\prime}} \|\mathbf{c}_{v^{\prime}}: v^{\prime} \in \mathcal{N}_{v}\right\}\right),
\end{equation}
where $\|$ denotes the concatenation operation. The context embedding is trained to represent the local structure of the node, which can help to distinguish the fraud. If we use addition operation for $\oplus$, then $\mathbf{h}_{v}\in \mathbb{R}^{2d}$.

\subsection{Neighbor Sampling}
\label{sec:neighbor sampling}
Since there exists a feature inconsistency problem, we should sample related neighbors rather than assign equal probabilities to them. Thus, we compute the \textbf{consistency score} between embeddings:
\begin{equation}
    s^{(l)}(u, v) = \exp\left(-\|\mathbf{h}_{u}^{(l)} - \mathbf{h}_{v}^{(l)}\|_{2}^{2}\right),
\end{equation}
where $s^{(l)}(\cdot,\cdot)$ denotes the consistency score for two nodes at $l$-th layer, and $\|\cdot\|_2$ is the $l_2$-norm\footnote{Other metrics, such as $l_1$-norm, are also applicable. } of vector. We first apply a threshold $\epsilon$ to filter neighbors far away from consistent. Then, we assign each node $u$ to the filtered neighbors $\Tilde{\mathcal{N}}_{v}$ of node $v$ with a sampling probability by normalizing its consistency score:
\begin{equation}~\label{eq:sampling prob}
     p^{(l)}(u; v) = s^{(l)}(u, v)/{\sum_{u\in\Tilde{\mathcal{N}}_{v}}s^{(l)}(u, v)}.
\end{equation}
Note that the probability is calculated at each layer for the $\operatorname{AGG}^{(l)}$.

\subsection{Relation Attention}
We have $R$ different relations in the graph. The relation information should also be included in the aggregation process to tackle the relation inconsistency problem. Hence, for each relation $r$, we train a relation vector $\mathbf{t}_r$, where $r=\{1,2,\dots,R\}$, to represent the relation information that should be incorporated. Since the relation information should be aggregated along with the neighbors to center node $v$, we adopt the self-attention mechanism~\cite{velivckovic2017graph} to assign weights for $Q$ sampled neighbor nodes:
\vspace{-2mm}
\begin{equation}
    \alpha_{q}^{(l)} = \exp\left(\sigma\left(\{\mathbf{h}_{q}^{(l)}\|\mathbf{t}_{r_q}\}\mathbf{a}^{\top}\right)\right)/{\sum_{q=1}^{Q}\exp\left(\sigma\left(\{\mathbf{h}_{q}^{(l)}\|\mathbf{t}_{r_q}\}\mathbf{a}^{\top}\right)\right)},
\end{equation}
where $r_q$ denotes the relation of $q$-th sample with node $v$, $\sigma$ is the activation function, and $a\in\mathbb{R}^{4d}$ represents the attention weights that is shared for all attention layer. The final $\operatorname{AGG}^{(l)}$ is:
\vspace{-2mm}
\begin{equation}
   \operatorname{AGG}^{(l)}\left(\left\{\mathbf{h}_{q}^{(l-1)}\right\}\Big|_{q=1}^{Q}\right) = \sum_{q=1}^{Q}\alpha_{q}^{(l)}\mathbf{h}_{q}^{(l)}, 
\end{equation}
where $\mathbf{h}_{q}^{(l)}$ is the embedding of $q$-th node sampled based on Eq.~(\ref{eq:sampling prob}). 

\section{EXPERIMENTS}

\subsection{Experimental Setup}

\subsubsection{Dataset and Graph Construction}
We utilize the YelpChi spam review dataset~\cite{Rayana2015}, along with three other benchmark datasets~\cite{kipf2016semi,hamilton2017inductive} to study the graph inconsistency problem in the fraud detection task.
The YelpChi spam review dataset includes hotel and restaurant reviews filtered (spam) and recommended (legitimate) by Yelp.
In this paper, we conduct a spam review classification task on the YelpChi dataset which is a binary classification problem.
We remove products with more than 800 reviews to restrict the size of the computation graph.
The pre-processed dataset has 29431 users, 182 products, and 45954 reviews (\%14.5 spams).


Based on previous studies~\cite{Rayana2015} which show the spam reviews have connections in user, product, rating, and time,
we take reviews as nodes in the graph and design three relations denoted by \textit{R-U-R}, \textit{R-S-R}, and \textit{R-T-R}. \textit{R-U-R} connects reviews posted by the same user; \textit{R-S-R} connects reviews under the same product with the same rating; \textit{R-T-R} connects two reviews under the same product posted in the same month.
We take the 100-dimension Word2Vec embedding of each review as its feature like previous work~\cite{li2019spam}.

\subsubsection{Baselines}
To show the ability of $\mathsf{GraphConsis}$ in alleviating inconsistency problems, we compare its performance with a non-GNN classifier, vanilla GNNs, and GNN-based fraud detectors.

\begin{itemize}
    \item \textbf{Logistic Regression}. A non-GNN classifier that makes predictions only based on the reviews features.
    
    \item \textbf{FdGars (GCN)}~\cite{wang2019fdgars}. A spam review detection algorithm using GCN~\cite{kipf2016semi}.
    
    \item \textbf{GraphSAGE}~\cite{hamilton2017inductive}. A popular GNN framework which samples neighboring nodes before aggregation.
    
    \item \textbf{Player2Vec}~\cite{zhang2019key}. A state-of-the-art fraud detection model which uses GCN to encode information in each relation, and uses GAT to aggregate neighbors from different relations.
\end{itemize}

\subsubsection{Experimental Settings}
We use Adam optimizer to train our model based on the cross-entropy loss. For the hyper-parameters, we choose $2$-layer structure, and the number of samples is set as $10$ and $5$ for the first layer and second layer, respectively. The embedding dimension of the hidden layer is $200$ and $100$ for the first layer and second layer, respectively.
We use F1-score to measure the overall classification performance and AUC to measure the performance of identifying spam reviews.

\subsection{The Inconsistency Problem}
We first take the Yelpchi dataset to demonstrate the inconsistency problem in applying GNN to fraud detection tasks. Table~\ref{tab:inconsistency} shows the statistics of graphs built on YelpChi comparing to node classification benchmark datasets used by~\cite{kipf2016semi,hamilton2017inductive}. \textit{Yelp-ALL} is composed of three single-relation graphs. 

Comparing to three widely-used benchmark node classification datasets, we find that a multi-relation graph constructed on YelpChi has a much higher density (the average node degree is greater than 100). It demonstrates that the real-world fraud graphs usually incorporate complex relations and neighbors, and thus render inconsistency problems. Before we compare the graph characteristics and analyze three inconsistency problems, similar to~\cite{hou2020measure}, we design two characteristic scores. One is the context characteristic score:
\begin{equation}\label{eq:context_chara}
    \gamma_{r}^{(c)} = \sum_{(u,v)\in{E}_{r}}\left(1-\mathbb{I}\left(u\sim v\right)\right)/|{E}_{r}|,
\end{equation}
where $\mathbb{I}(\cdot)\in\{0,1\}$ is an indicator function to indicate whether node $u$ and node $v$ have the same label. We sum all the indication w.r.t. all the edges and normalized by the total number of edges $|{E}_{r}|$. The context characteristic measures the label similarity between neighboring nodes under a specific relation $r$. The other one is the feature characteristic score:
\begin{equation}\label{eq:feature_chara}
    \gamma_{r}^{(f)}=\sum_{(u,v) \in{E}_{r}}\exp\left(-\left\|\mathbf{x}_{u}-\mathbf{x}_{v}\right\|^{2}_{2}\right)/{|{E}_{r}|\cdot d},
\end{equation}
where we employ the RBF kernel function\footnote{Other kernel functions can also be applied.} as the similarity measurement between two connected nodes. The overall feature characteristic score is normalized by the product of the total number of edges $|E_{r}|$ and the feature dimension $d$. Normalizing the similarity by feature dimension is to fairly compare the feature characteristics of different graphs, which may have different feature dimensions.

\vspace{1mm}

\noindent \textbf{Context Inconsistency.} We compute the context characteristic $\gamma_{r}^{(c)}$ based on Eq.~(\ref{eq:context_chara}), which measures the context consistency. For the graph \textit{R-T-R}, \textit{R-S-R} and \textit{Yelp-ALL}, there are less than $10\%$ of neighboring nodes have similar labels. It shows that fraudsters may hide themselves among regular entities under some relations. 

\vspace{1mm}

\noindent \textbf{Feature Inconsistency.} We calculate the feature characteristic $\gamma_{r}^{(f)}$ using Eq.~(\ref{eq:feature_chara}). The graph constructed by \textit{R-U-R} relation~(reviews posted by the same user) has higher feature characteristic than the other two relations. Thus, we need to sample the neighboring nodes not only based on their relations but also the feature similarities.

\vspace{1mm}

\noindent \textbf{Relation Inconsistency.} For graphs constructed by three different relations, the neighboring nodes also have different feature/label inconsistency score. Thus, we need to treat different relations with different attention weights during the aggregation. 

\vspace{-3mm}
\begin{table}[h]
\small
\centering
\caption{The statistics of different graphs.}
\resizebox{0.75\linewidth}{!}{%
\begin{tabular}{c|lcccc}  
\hline
\multicolumn{2}{c}{\textbf{Graph}} & \textbf{\#Nodes}  & \textbf{\#Edges}  & \textbf{$\gamma^{(f)}$}  & \textbf{$\gamma^{(c)}$}  \\ 
\hline
\multirow{3}{*}{\rotatebox[origin=c]{90}{\textbf{Others}}} & \textbf{Cora} & 2,708  & 5,278    & 0.72 & 0.81 \\
& \textbf{PPI} & 14,755  & 225,270 & 0.48 & 0.98\\
& \textbf{Reddit} &  232,965  & 11,606,919   & 0.70 & 0.63 \\
\hline
 \multirow{4}{*}{\rotatebox[origin=c]{90}{\textbf{Ours}}} & \textbf{\textit{R-U-R}} & 45,954 & 98,630   & 0.83 & 0.90 \\
& \textbf{\textit{R-T-R}} & 45,954 & 1,147,232   &0.79 & 0.05\\
& \textbf{\textit{R-S-R}} & 45,954 & 6,805,486  & 0.77 & 0.05\\
& \textbf{\textit{Yelp-ALL}} & 45,954 & 7,693,958   & 0.77& 0.07\\
\hline
\end{tabular}}
\label{tab:inconsistency}
\end{table} 
\vspace{-3mm}



\subsection{Performance Evaluation}
Table~\ref{tab:exp_result} shows the experiment results of the spam review detection task. We could see that $\mathsf{GraphConsis}$ outperforms other models under $80\%$ and $60\%$ of training data on both metrics, which suggests that we can alleviate the inconsistency problem. Compared with other GNN-based models, LR performs stably and better on AUC. It indicates that the node feature is useful, but the aggregator in GNN undermines the classifier in identifying fraudsters. This observation also proves that the inconsistency problem is critical and should be considered when applying GNNs to fraud detection tasks. Compared to Player2Vec which also learns relation attention, $\mathsf{GraphConsis}$ performs better. It suggests that solely using relation attention cannot alleviate the feature inconsistency. The neighbors should be filtered and then sampled based on our designed methods. FdGars directly aggregates neighbors' information and GraphSAGE samples neighbors with equal probability. Both of them perform worse than $\mathsf{GraphConsis}$, which shows that our neighbor sampling techniques are useful.

\begin{table}[h]
\small
\centering
\caption{Experiment results under different training \%.}
\resizebox{1\linewidth}{!}{%
\begin{tabular}{ccccccc}  
\hline
\multirow{2}{*}{\textbf{Method}} & \multicolumn{2}{c}{\textbf{40\%}} & \multicolumn{2}{c}{\textbf{60\%}}& \multicolumn{2}{c}{\textbf{80\%}} \\ 
\cline{2-7}
 & F1 & AUC & F1 & AUC & F1 & AUC\\
\hline
\textbf{LR} & 0.4647  & \textbf{0.6140}  & 0.4640  & 0.6239 & 0.4644 & 0.6746 \\
\textbf{GraphSAGE} & 0.4956  & 0.5081  & 0.5127  & 0.5165 & 0.5158 & 0.5169 \\
\textbf{FdGars} &  0.4603  & 0.5505  & 0.4600  & 0.5468 & 0.4603 & 0.5470 \\
\textbf{Player2Vec} & 0.4608 & 0.5426  & 0.4608  & 0.5697 &0.4608  & 0.5403\\
\textbf{GraphConsis} & \textbf{0.5656} & 0.5911  & \textbf{0.5888}  & \textbf{0.6613} & \textbf{0.5776}  & \textbf{0.7428}\\
\hline
\end{tabular}}
\label{tab:exp_result}
\end{table} 

\section{Conclusion and Future Works}
In this paper, we investigate three inconsistency problems in applying GNNs in fraud detection problem.
To address those problems, we design three modules respectively and propose $\mathsf{GraphConsis}$. Experiment results show the effectiveness of $\mathsf{GraphConsis}$. Future work includes devising an adaptive sampling threshold for each relation to maximize the receptive field of GNNs. Investigating the inconsistency problems under other fraud datasets is another avenue of future research.

\begin{acks}
This work is supported by the National Key R\&D Program of China under grant 2018YFC0830804, and in part by NSF under grants III-1526499, III-1763325, III-1909323, and CNS-1930941. For any correspondence, please refer to Hao Peng.
\end{acks}

\bibliographystyle{ACM-Reference-Format}
\bibliography{sample-base}


\end{document}